\documentclass[]{article}
\usepackage{times}
\usepackage{cite}
\usepackage{psfrag}
\usepackage[T1]{fontenc}
\usepackage[latin9]{inputenc}
\usepackage{geometry}

\usepackage{float}
\usepackage{amsmath}
\usepackage{amssymb}
\usepackage{graphicx}

\makeatletter

\providecommand{\tabularnewline}{\\}

\usepackage{dsfont}
\date{}

\makeatother

\begin{document}

%




\title{Band gaps in the relaxed linear micromorphic continuum}

\author{A. Madeo$^{1,6}$\footnote{Corresponding author,~e-mail:~\textsf{angela.madeo@insa-lyon.fr}}, P. Neff$^{2,6}$, I.-D. Ghiba$^{2,3,6}$, L. Placidi$^{4,6}$ and G. Rosi$^{5,6}$.}%
 \maketitle

\begin{enumerate}
\item Université de Lyon-INSA, 20 Av. Albert Einstein, 69100
Villeurbanne Cedex, France. 
\item Lehrstuhl f\"{u}r Nichtlineare Analysis und Modellierung,
Fakultät für Mathematik, Universität Duisburg-Essen, Campus Essen,
Thea-Leymann Str. 9, 45127 Essen, Germany. 
\item Dep. of Mathematics, University \textquotedblleft{}A.I.
Cuza\textquotedblright{}, 11 Blvd. Carol I, I\c{a}si; Octav Mayer
Institute of Mathematics, Romanian Academy, I\c{a}si; Institute of
Solid Mechanics, Romanian Academy, Bucharest, Romania. 
\item Università Telematica Internazionale Uninettuno, Corso
V. Emanuele II 39, 00186 Roma, Italy. 
\item Laboratoire Modélisation Multi Echelle, MSME UMR 8208 CNRS, Université
Paris-Est, 61 Avenue du Géneral de Gaulle, Créteil Cedex 94010, France. 
\item International Research Center for the Mathematics and Mechanics of Complex Systems (M\&MoCS), Universit\`{a} dell'Aquila, Palazzo Caetani, 04012 Cisterna di Latina, Italia 
\end{enumerate}

\begin{abstract}
In this note we show that the relaxed linear micromorphic model recently
proposed by the authors can be suitably used to describe the presence
of band-gaps in metamaterials with microstructures in which strong
contrasts of the mechanical properties are present (e.g. phononic
crystals and lattice structures). This relaxed micromorphic model
only has 6 constitutive parameters instead of 18 parameters needed
in Mindlin- and Eringen-type classical micromorphic models. We show
that the onset of band-gaps is related to a unique constitutive parameter,
the \textit{Cosserat couple modulus} $\mu_{c}$ which starts to account
for band-gaps when reaching a suitable threshold value. The limited
number of parameters of our model, as well as the specific effect
of some of them on wave propagation can be seen as an important step
towards indirect measurement campaigns.
\end{abstract}

\section{Introduction}

Micromorphic models were originally proposed by Mindlin \cite{Mindlin}
and Eringen \cite{EringenBook} in order to study materials with microstructure
while remaining in the simplified framework of macroscopic continuum
theories. Nevertheless, the huge number of material parameters (18
in the linear-isotropic case) limited up to now the application of
these powerful theories to describe the behavior of real metamaterials
(see e.g. \cite{Neff_Forest_jel05}). In this paper, we propose to
use the newly developed relaxed micromorphic model presented in \cite{Ghiba,NeffUnif}
to study wave propagation in microstructured materials which exhibit
frequency band-gaps. In the present paper we want to focus on a new
and fascinating development in generalized continuum models applied
to band-gaps wave phenomena. In a companion paper \cite{madeo2013wave}
we have detailed how the proposed relaxed model can be compared with
classical micromorphic and with second gradient and Cosserat models.
Here, we mainly summarize our results concerning the description of
frequency band-gaps in relaxed micromorphic continua and we detail
how the effect of the Cosserat couple modulus $\mu_{c}$ is essential
for their description.

The proposed relaxed model only counts six constitutive parameters and
is fully able to account for the effect of microstructure on the macroscopic
mechanical behavior of the considered media. The request of positive-definiteness
for the proposed relaxed model is weaker with respect to the classical
Mindlin-Eringen model (positive-definiteness of the energy with the
whole gradient term). This weaker request allows the relaxed model
to account for weaker connections at the microscopic level compared
to those which are possible in classical micromorphic theories. For
this reason the proposed relaxed model allows the description of band
gaps while the classical approach does not.

It is known that some materials like phononic crystals and lattice
structures (see e.g. \cite{Vasseur1}), granular assemblies with defects
(see e.g. \cite{Kafesaki,Merkel3,Merkel2,MerkelGranular}) and composites
(\cite{Economou,Vasseur}) can inhibit wave propagation in particular
frequency ranges (band-gaps). The aim of this note is to show that
the proposed relaxed model allows for describing frequency band-gaps
by ``switching on'' a unique constitutive parameter which is known
as \textit{Cosserat couple modulus} $\mu_{c}$ (see e.g. \cite{Neff_JeongMMS08,Jeong_Neff_ZAMM08,Neff_ZAMM05,Neff_Jeong_Conformal_ZAMM08,Neff_Jeong_bounded_stiffness09}).
The limited number of constitutive parameters makes possible the future
conception of direct and indirect measurements on real materials exhibiting
frequency band-gaps. On the other hand, the generality of the proposed
relaxed model can also be seen as a tool to aid the engineering design
of new metamaterials with improved band-gap capabilities. Materials
of this type could be used as an alternative to piezoelectric materials
which are used today for vibration control and which are for this
reason extensively studied in the literature (see e.g. \cite{Piezo1,isola2,isola6,isola1,isola3,Piezo}).
Because of the possible interest of our findings in a linear modelling
framework, we summarize in this communication the main novel results
on band gaps related to our relaxed micromorphic model.

\section{Equations of motion in strong form}

As shown in \cite{NeffUnif} the equations of motion of the considered
linear relaxed isotropic and homogeneous micromorphic continuum, when neglecting body loads, read

\begin{align}
\rho\,\mathbf{u}_{tt}=& \mathrm{Div}\left[2\,\mu_{e}\,\mathrm{sym}\left(\nabla\mathbf{u}-\mathbf{P}\right)+\lambda_{e}\mathrm{tr}\left(\nabla\mathbf{u}-\mathbf{P}\right)\mathds1+2\,\mu_{c}\,\mathrm{skew}\left(\nabla\mathbf{u}-\mathbf{P}\right)\right]\nonumber  \\
\label{eq:bulk-mod-3}\\
\eta\,\mathbf{P}_{tt}=&-\alpha_{c}\,\mathrm{\mathrm{Cur}l}\left(\mathrm{\mathrm{Cur}l}\,\mathbf{P}\right)+  2\,\mu_{e}\,\mathrm{sym}\left(\nabla\mathbf{u}-\mathbf{P}\right)+\lambda_{e}\mathrm{tr}\left(\nabla\mathbf{u}-\mathbf{P}\right)\mathds1\nonumber \\
&-2\,\mu_{h}\,\mathrm{sym}\,\mathbf{P}-\lambda_{h}\mathrm{tr}\,\mathbf{P}\:\mathds1+2\,\mu_{c}\,\mathrm{skew}\left(\nabla\mathbf{u}-\mathbf{P}\right),\nonumber
\end{align}
where $\mathbf{u}\in\mathbb{R}^{3}$ is the displacement field, $\mathbf{P}\in\mathbb{R}^{3\times3}$
is the microdeformation tensor (basic kinematical fields), $\rho$
and $\eta$ are the macro and micro mass densities respectively and
all other quantities are the constitutive parameters of the model.
As for the operators appearing in (\ref{eq:bulk-mod-3}), we use the
following notations
\[
\mathrm{sym}\,\mathbf{X}=\frac{\mathbf{X}+\mathbf{X}^{T}}{2}\qquad\mathrm{skew}\,\mathbf{X}=\frac{\mathbf{X}-\mathbf{X}^{T}}{2},\qquad\mathrm{tr}\,\mathbf{X}=X_{ii},
\]
\[
(\mathrm{Div}\,\mathbf{X})_{i}=X_{ij,j},\qquad\left(\nabla\mathbf{u}\right)_{ij}=u_{i,j},\qquad(\mathrm{Curl}\:\mathbf{X}_{ij}=X_{ia,b}\epsilon_{jab}),
\]
where $\epsilon_{jab}$ is the classical Levi-Civita symbol, $X_{ij}$
denote the components of the tensor $\mathbf{X}$, the subscript $t$ indicates a time deivative, the subscript $j$
indicates the derivative with respect to the $j$-th component of
the space variable and the Einstein summation convention over repeated
indices has been adopted. It can be checked that when considering
a completely one-dimensional case, the term $\mathrm{\mathrm{Cur}l}\left(\mathrm{\mathrm{Cur}l}\,\mathbf{P}\right)$
vanishes and no characteristic length related to microstructure can
be accounted for by our model. We need at least the case of plane
waves (all the components of $\mathbf{u}$ and $\mathbf{P}$ are non-vanishing,
but all depend on one space variable $X$ which is also the direction
of propagation of considered wave) to disclose all the characteristic
features of the proposed relaxed model. On the other hand, the Mindlin-Eringen
models allow to account for characteristic lengths even when considering
complete one-dimensional cases (all components of the kinematical
fields in the plane orthogonal to the direction of propagation are
vanishing). This is shown e.g. in \cite{MauginBerezowski,MauginBookWaves}
in which these fully 1D equations are derived by the standard internal
variable theory. We also remark that, in general, the relaxed term
$\mathrm{\mathrm{Cur}l}\left(\mathrm{\mathrm{Cur}l}\,\mathbf{P}\right)$
in the second of Eqs. (\ref{eq:bulk-mod-3}) is much weaker than the
full term $\Delta\mathbf{P}$ appearing in Mindlin- and Eringen-type
models. Despite this weaker formulation, we claim that the proposed
relaxed model is fully able to account for the presence of microstructure
on the overall mechanical behavior of the considered continua. In
particular, our relaxed model is able to account for the description
of frequency band-gaps, while the classical Mindlin- and Eringen-type
models are not.

In \cite{NeffUnif} it is also proved that positive definiteness of
the strain energy density associated to Eqs.(\ref{eq:bulk-mod-3})
implies
\begin{align}
\mu_{e}>0,\quad\mu_{c}\geq0,\quad3\lambda_{e}+2\mu_{e}>0,\quad\mu_{h}>0,\quad 3\lambda_{h}+2\mu_{h}>0,\quad\alpha_{c}>0\label{DefPos}
\end{align}
In this note we will only consider non-vanishing, positive values
of the Cosserat couple modulus $\mu_c$ which is seen as a trigger of the band-gap
phenomenon.

We limit ourselves to the case of plane waves travelling in an infinite
domain, i.e. we suppose that the space dependence of all the introduced
kinematical fields is limited only to the space component $X$ which
we also suppose to be the direction of propagation of the considered
wave. We introduce the new variables
\begin{gather}
P^{S}:=\frac{1}{3}\left(P_{11}+P_{22}+P_{33}\right),\quad P^{D}:=\left(\mathrm{dev\: sym}\,\mathbf{P}\right)_{11},\nonumber \\
\label{SpherDev}\\
P_{(1\xi)}:=\left(\mathrm{sym}\:\mathbf{P}\right)_{1\xi},
\quad P_{\left[1\xi\right]}:=\left(\mathrm{skew}\:\mathbf{P}\right)_{1\xi},\quad\xi=1,2.\nonumber
\end{gather}
It is immediate that, according to the Cartan-Lie-algebra decomposition
(see e.g. \cite{NeffUnif}), the component $P_{11}$ of the tensor
$\mathbf{P}$ can be rewritten as $P_{11}=P^{D}+P^{S}$. We also define
the additional variables
\begin{equation}
P^{V}=P_{22}-P_{33},\qquad P_{(1\xi)}=\left(\mathrm{sym}\:\mathbf{P}\right)_{1\xi},\qquad P_{\left[1\xi\right]}=\left(\mathrm{skew}\:\mathbf{P}\right)_{1\xi},\quad\xi=1,2.\label{SymSkew}
\end{equation}

We rewrite the equations of motion (\ref{eq:bulk-mod-3}) in terms
of the new variables (\ref{SpherDev}), (\ref{SymSkew}) and, of course,
of the displacement variables. Before doing so, we introduce the quantities%
\footnote{Due to the chosen values of the parameters, which are supposed to
satisfy (\ref{DefPos}), all the introduced characteristic velocities
and frequencies are real. Indeed, the condition $3\lambda_{e}+2\mu_{e}>0$,
together with the condition $\mu_{e}>0$, imply the condition $\lambda_{e}+2\mu_{e}>0$. %
}
\begin{gather}
c_{m}=\sqrt{\frac{\alpha_{c}}{\eta}},\qquad c_{s}=\sqrt{\frac{\mu_{e}+\mu_{c}}{\rho}},\qquad c_{p}=\sqrt{\frac{\lambda_{e}+2\mu_{e}}{\rho}},\nonumber\\
\omega_{s}=\sqrt{\frac{2\left(\mu_{e}+\mu_{h}\right)}{\eta}},\quad\omega_{p}=\sqrt{\frac{\left(3\lambda_{e}+2\mu_{e}\right)+\left(3\lambda_{h}+2\mu_{h}\right)}{\eta}},\quad\omega_{r}=\sqrt{\frac{2\mu_{c}}{\eta}},\label{Definitions}\\
\omega_{l}=\sqrt{\frac{\lambda_{h}+2\mu_{h}}{\eta}},\quad\omega_{t}=\sqrt{\frac{\mu_{h}}{\eta}}.\nonumber
\end{gather}
With the proposed new choice of variables and recalling that we are
considering the case of planar waves, we are able to rewrite the governing
equations (\ref{eq:bulk-mod-3}) as different uncoupled sets of equations,
namely:
\begin{itemize}
\item A set of three equations only involving longitudinal quantities (left)
and two sets of three equations only involving transverse quantities
in the $k$-th direction:
\end{itemize}
\begin{align}
&\begin{cases}
\ddot{u}_{1}=c_{p}^{2}u_{1,11}-\frac{2\mu_{e}}{\rho}\, P_{,1}^{D}-\frac{3\lambda_{e}+2\mu_{e}}{\rho}\, P_{,1}^{S},\\
\\
\ddot{P}^{D}=\frac{4}{3}\,\frac{\mu_{e}}{\eta}u_{1,1}+\frac{1}{3}c_{m}^{2}P_{,11}^{D}-\frac{2}{3}c_{m}^{2}P_{,11}^{S}-\omega_{s}^{2}P^{D},\\
\\
\ddot{P}^{S}=\frac{3\lambda_{e}+2\mu_{e}}{3\eta}u_{1,1}-\frac{1}{3}c_{m}^{2}P_{,11}^{D}+\frac{2}{3}c_{m}^{2}P_{,11}^{S}-\omega_{p}^{2}P^{S},
\end{cases} \nonumber\\
\label{Longitudinal}\\
&\begin{cases}
\ddot{u}_{\xi}=c_{s}^{2}u_{\xi,11}-\frac{2\mu_{e}}{\rho}\, P_{\left(1\xi\right),1}+\frac{\eta}{\rho}\omega_{r}^{2}P_{\left[1\xi\right],1},\\
\\
\ddot{P}_{\left(1\xi\right)}=\frac{\mu_{e}}{\eta}u_{\xi,1}+\frac{1}{2}c_{m}^{2}P_{(1\xi)}{}_{,11}+\frac{1}{2}c_{m}^{2}P_{\left[1\xi\right],11}-\omega_{s}^{2}P_{(1\xi)},\\
\\
\ddot{P}_{\left[1\xi\right]}=-\frac{1}{2}\omega_{r}^{2}u_{\xi,1}+\frac{1}{2}c_{m}^{2}P_{(1\xi),11}+\frac{1}{2}c_{m}^{2}P_{\left[1\xi\right]}{}_{,11}-\omega_{r}^{2}P_{\left[1\xi\right]},\nonumber
\end{cases}
\end{align}
, with $\xi=2,3$.
\begin{itemize}
\item Three uncoupled equations only involving the variables $P_{\left(23\right)}$,
$P_{\left[23\right]}$ and $P^{V}$ respectively
\end{itemize}
\begin{align}
\ddot{P}_{\left(23\right)} =-\omega_{s}^{2}P_{\left(23\right)}+c_{m}^{2}P_{\left(23\right),11},\quad\ddot{P}_{\left[23\right]}=-\omega_{r}^{2}P_{\left[23\right]}+c_{m}^{2}P_{\left[23\right],11},\nonumber\\
\label{Shear}\\
\ddot{P}^{V}=-\omega_{s}^{2}P^{V}+c_{m}^{2}P_{,11}^{V}.\nonumber
\end{align}

These 12 scalar differential equations will be used to study wave
propagation in our relaxed micromorphic media.

\section{Plane wave propagation}

We now look for a wave form solution of the previously derived equations
of motion. We start from the uncoupled Eqs. (\ref{Shear}) and
assume that the involved unknown variables take the harmonic form
\begin{align}
&P_{\left(23\right)}=Re\left\{ \beta_{\left(23\right)}e^{i(kX-\omega t)}\right\} ,\quad P_{\left[23\right]}=Re\left\{ \beta_{\left[23\right]}e^{i(kX-\omega t)}\right\} ,\nonumber\\
\label{WaveForm1}\\
&P^{V}=Re\left\{ \beta^{V}e^{i(kX-\omega t)}\right\} \nonumber,
\end{align}
where $\beta_{\left(23\right)}$, $\beta_{\left[23\right]}$ and $\beta^{V}$
are the amplitudes of the three introduced waves. Replacing this wave
form in Eqs. (\ref{Shear}) and simplifying one obtains the following
dispersion relations respectively:
\begin{equation}
\omega(k)=\sqrt{\omega_{s}^{2}+k^{2}c_{m}^{2}},\qquad\omega(k)=\sqrt{\omega_{r}^{2}+k^{2}c_{m}^{2}},\qquad\omega(k)=\sqrt{\omega_{s}^{2}+k^{2}c_{m}^{2}}.\label{Eigen1}
\end{equation}
We notice that for a vanishing wave number ($k=0$) the dispersion
relations for the three considered waves give non-vanishing frequencies
so that these waves are so-called \textit{optic waves} with cutoff
frequencies $\omega_{s}$, $\omega_{r}$ and $\omega_{s}$ respectively.

We now introduce the unknown vectors $\mathbf{v}_{1}=\left(u_{1},P^{D},P^{S}\right)$
and $\mathbf{v}_{\xi}=\left(u_{\xi},P_{(1\xi)},P_{[1\xi]}\right),\ \xi=2,3$
and look for wave form solutions of equations (\ref{Longitudinal})
in the form
\begin{equation}
\mathbf{v}_{1}=Re\left\{ \boldsymbol{\beta}e^{i(kX-\omega t)}\right\} ,\qquad\mathbf{v}_{\xi}=Re\left\{ \boldsymbol{\gamma}^{\xi}e^{i(kX-\omega t)}\right\} ,\ \xi=2,3,\label{WaveForm2}
\end{equation}
where $\boldsymbol{\beta}=(\beta_{1},\beta_{2},\beta_{3})^{T}$ and
$\boldsymbol{\gamma}^{\xi}=(\gamma_{1}^{\xi},\gamma_{2}^{\xi},\gamma_{3}^{\xi})^{T}$
are the unknown amplitudes of the considered waves. Replacing this
expressions in Eqs. (\ref{Longitudinal}) one gets respectively
\begin{equation}
\mathbf{A}_{1}\cdot\boldsymbol{\beta}=0,\qquad\qquad\mathbf{A}_{\xi}\cdot\boldsymbol{\gamma}^{\xi}=0,\qquad\qquad\xi=2,3,\label{AlgSys}
\end{equation}
where
\begin{gather*}
\mathbf{A}_{1}=\left(\begin{array}{ccc}
-\omega^{2}+c_{p}^{2}\, k^{2} & \, i\: k\:2\mu_{e}/\rho\  & i\: k\:\left(3\lambda_{e}+2\mu_{e}\right)/\rho\\
\\
-i\: k\,\frac{4}{3}\,\mu_{e}/\eta & -\omega^{2}+\frac{1}{3}k^{2}c_{m}^{2}+\omega_{s}^{2} & -\frac{2}{3}\, k^{2}c_{m}^{2}\\
\\
-\frac{1}{3}\, i\, k\:\left(3\lambda_{e}+2\mu_{e}\right)/\eta & -\frac{1}{3}\, k^{2}\, c_{m}^{2} & -\omega^{2}+\frac{2}{3}\, k^{2}\, c_{m}^{2}+\omega_{p}^{2}
\end{array}\right),\\
\\
\\
\mathbf{A}_{2}=\mathbf{A}_{3}=\left(\begin{array}{ccc}
-\omega^{2}+k^{2}c_{s}^{2}\  & \, i\, k\,2\mu_{e}/\rho\  & -i\, k\,\frac{\eta}{\rho}\omega_{r}^{2},\\
\\
-\, i\, k\,2\mu_{e}/\eta, & -2\omega^{2}+k^{2}c_{m}^{2}+2\omega_{s}^{2} & k^{2}c_{m}^{2}\\
\\
i\, k\,\omega_{r}^{2} & k^{2}c_{m}^{2} & -2\omega^{2}+k^{2}c_{m}^{2}+2\omega_{r}^{2}
\end{array}\right).
\end{gather*}
In order to have non-trivial solutions of the algebraic systems (\ref{AlgSys}),
one must impose that
\begin{equation}
\mathrm{det}\,\mathbf{A}_{1}=0,\qquad\qquad\mathrm{det}\,\mathbf{A}_{2}=0,\qquad\qquad\mathrm{det}\,\mathbf{A}_{3}=0,\label{Dispersion}
\end{equation}
which are the so-called dispersion relations $\omega=\omega\left(k\right)$
for the longitudinal and transverse waves in the relaxed micromorphic
continuum.

\section{\label{NumericalSim}Numerical results}

In this section, following Mindlin \cite{Mindlin,EringenBook}, we
will show the dispersion relations $\omega=\omega(k)$ associated
to our relaxed micromorphic model.
\begin{table}[H]
\begin{centering}
\begin{tabular}{cccc}
\hline
Parameter & Value & Unit & Description  \tabularnewline
\hline
\hline
$\lambda$ & $82.5$ & $MPa$ & Macroscopic Lamé modulus\tabularnewline
\hline
$\mu$ & $66.7$ & $MPa$ & Macroscopic Lamé modulus\tabularnewline
\hline
$\rho$ & $2500$ & $Kg/m^{3}$ & Macroscopic mass density\tabularnewline
\hline
$\rho'$ & $2500$ & $Kg/m^{3}$ & Microscopic mass density\tabularnewline
\hline
\hline
$\mu_{e}$ & $200$ & $MPa$ & Isotropic scale transition parameter\tabularnewline
\hline
$\lambda_{e}=2\mu_{e}$ & 400 & $MPa$ & Isotropic scale transition parameter \tabularnewline
\hline
\hline
$\lambda_{h}$ & $100$ & $MPa$ & Microscopic Lamé modulus \tabularnewline
\hline
$\mu_{h}$ & 100 & $MPa$ & Microscopic Lamé modulus  \tabularnewline
\hline
$d$ & $2$ & $mm$ & Characteristic size of inclusions \tabularnewline
\hline
$L_{c}$  & $3$ & $mm$  & Internal length \tabularnewline
\hline
$\eta=d^{2}\,\rho'$ & $10^{-2}$ & $Kg/m$  & Micro-inertia\tabularnewline
\hline
$\alpha_{c}=\mu_{e}L_{c}^{2}$ & $1.8\times10^{-3}$ & $MPa\: m^{2}$ & Microscopic stiffness  \tabularnewline
\hline

\end{tabular}
\par\end{centering}

\caption{\label{ParametersValues}Values of the parameters of the relaxed model
used in the numerical simulations.}

\end{table}
 We start by showing in Tab. \ref{ParametersValues}
the values of the parameters of the relaxed model used in the performed
numerical simulations. In order to make the obtained results better
exploitable, we also recall that in \cite{NeffUnif,Neff_Forest_jel05}
the following homogenized formulas were obtained which relate the
parameters of the relaxed model to the macroscopic Lamé parameters
$\lambda$ and $\mu$ which are usually measured by means of standard
mechanical tests
\begin{equation}
\mu_{e}=\frac{\mu_{h}\,\mu}{\mu_{h}-\mu},\qquad2\mu_{e}+3\lambda_{e}=\frac{(2\mu_{h}+3\lambda_{h})\left(2\mu+3\lambda\right)}{(2\mu_{h}+3\lambda_{h})-\left(2\mu+3\lambda\right)}.\label{Homogenized}
\end{equation}
These relationships imply that the following inequalities are satisfied:
$\mu_{h}>\mu$, $3\lambda_{h}+2\mu_{h}>3\lambda+2\mu$. It is clear
that, once the values of the parameters of the relaxed models are
known, the standard Lamé parameters can be calculated by means of
formulas (\ref{Homogenized}), which is what was done in Tab. \ref{ParametersValues}.
The table also includes the values of
two characteristic lengths $L_{c}$ and $d$ which can be respectively
associated to the characteristic length corresponding to the relaxed
micromorphic coefficient $\alpha_{c}$ and to the characteristic size
of microscopic inclusions. Indeed, while the parameter $d$ can be
directly associated to the size of microstructure (see also \cite{Mindlin}),
the physical meaning of the parameter $L_{c}$ is more difficult to
be visualized. This latter can be indeed associated to the characteristic
size of the region of interactions of a Representative Elementary
Volume with the surrounding ones. The characteristic length $L_{c}$
can be hence associated, for example, to the size of some particular
boundary layers which can be observed in the deformation patterns
of micro-structured materials subjected to particular types of loading
and/or boundary conditions. Finally, it should be noted that the micro inertia $\eta$ is calculated using the microscopic mass density $\rho$, which is the density of inclusions, as suggested by Mindlin \cite{Mindlin63}.

The numerical values of the parameters listed in Tab. \ref{ParametersValues}
have been chosen in the set of all possible values in order to let
band-gaps clearly appear in a frequency range which can be considered
to be of physical interest. Indeed, the band gaps which are shown
in this paper range between $80$ and $200\, kHz$, which are values
proposed for band gaps in phononic crystals (see e.g.\cite{FrequencyOfBandGaps}).
The Young's modulus of the homogenized material considered in the
present paper corresponds to the one of a rather soft material. Nevertheless,
stiffer materials can also be considered which still show frequency
band-gaps by slightly changing the other involved parameters.

It is evident (see Eqs. (\ref{Definitions})) that, in general, the
relative positions of the horizontal asymptotes $\omega_{l}$ and
$\omega_{t}$ as well as of the cutoff frequencies $\omega_{s}$,
$\omega_{r}$ and $\omega_{p}$ can vary depending on the values of
the constitutive parameters (see (\ref{Definitions})). It can also
be checked that, in the case in which $\lambda_{e}>0$ and $\lambda_{h}>0$
one always has $\omega_{p}>\omega_{s}>\omega_{t}$ and $\omega_{l}>\omega_{t}$:
we will keep this hypothesis in the remainder of the paper. The relative
position of $\omega_{l}$ and of $\omega_{s}$ can vary depending
on the values of the parameters $\lambda_{h}$ and $\mu_{h}$. It
can be finally recognized that, in order to have a global band-gap,
i.e. a frequency range in which no one of the considered types of
waves can propagate, the following conditions must be simultaneously
satisfied: $\omega_{s}>\omega_{l}$ and $\omega_{r}>\omega_{l}$.
In terms of the constitutive parameters of the relaxed model, we can
say that global band-gaps can exist, in the case in which one considers
positive values for the parameters $\lambda_{e}$ and $\lambda_{h}$
, if and only if we have simultaneously
\begin{gather}
0<\mu_{e}<+\infty,\qquad\qquad0<\lambda_{h}<2\mu_{e},\qquad\qquad\mu_{c}>\frac{\lambda_{h}+2\text{\ensuremath{\mu}}_{h}}{2}=:\mu_{c}^{0}.\label{Band Gaps}
\end{gather}
As far as negative values for $\lambda_{e}$ and $\lambda_{h}$ are
allowed, the conditions for band gaps are not so straightforward as
(\ref{Band Gaps}), but we do not consider this possibility in this
note. In the numerical simulations presented in this note we choose
to test three characteristic values of the Cosserat couple modulus
as listed in Tab. \ref{tab:muc}.
\begin{table}[h]
\begin{centering}
\begin{tabular}{ccc}
\hline
Case & Value & Unit\tabularnewline
\hline
1 &  $\mu_{c}=\mu_{c}^{0}=1.5$ & $MPa$\tabularnewline
\hline
2 &  $\mu_{c}=2\mu_{c}^{0}=3$ & $MPa$\tabularnewline
\hline
3 &  $\mu_{c}=3\mu_{c}^{0}=4.5$ & $MPa$\tabularnewline
\hline
\end{tabular}
\par\end{centering}

\caption{\label{tab:muc}Characteristic values of the Cosserat couple modulus
$\mu_{c}$ chosen for the numerical simulations (the values of the
parameters $\lambda_{h}$ et $\mu_{h}$ are given in Tab. \ref{ParametersValues}).}

\end{table}
 The choice of these characteristic values allows to show how the
Cosserat couple modulus is indeed the parameter which is responsible
for the onset of band-gaps in the considered generalized continuum.

\begin{figure}[h]
\begin{centering}
\begin{tabular}{ccccc}
\includegraphics[width=0.3\linewidth]{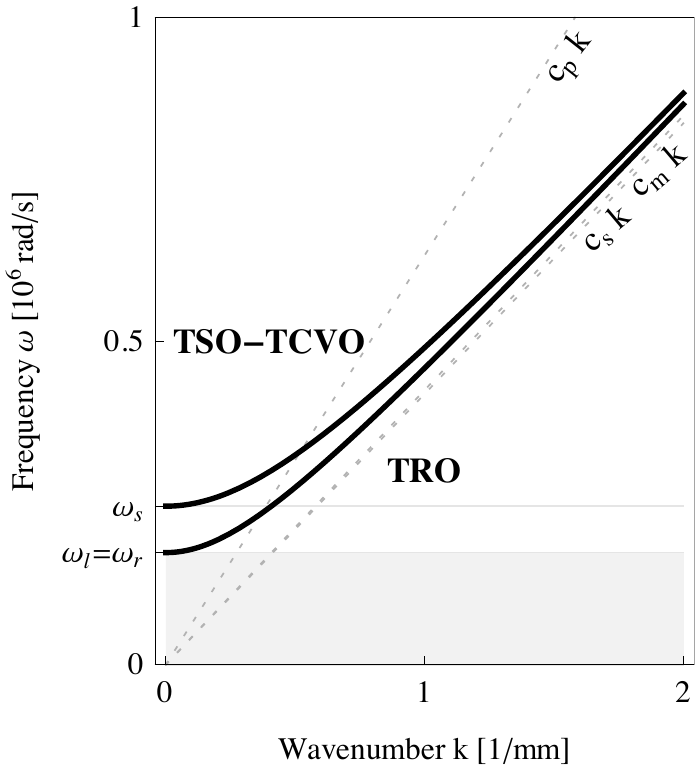} & \includegraphics[width=0.3\linewidth]{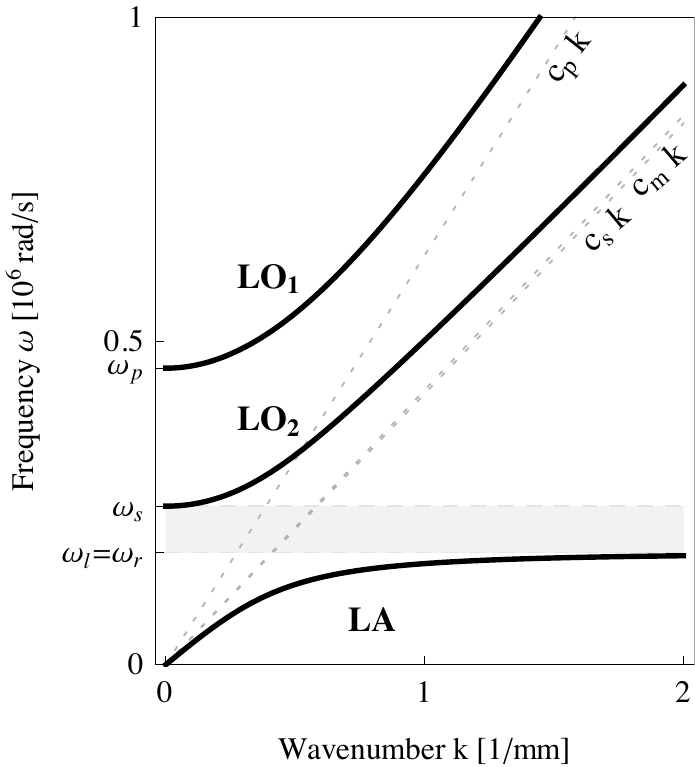} & \includegraphics[width=0.3\linewidth]{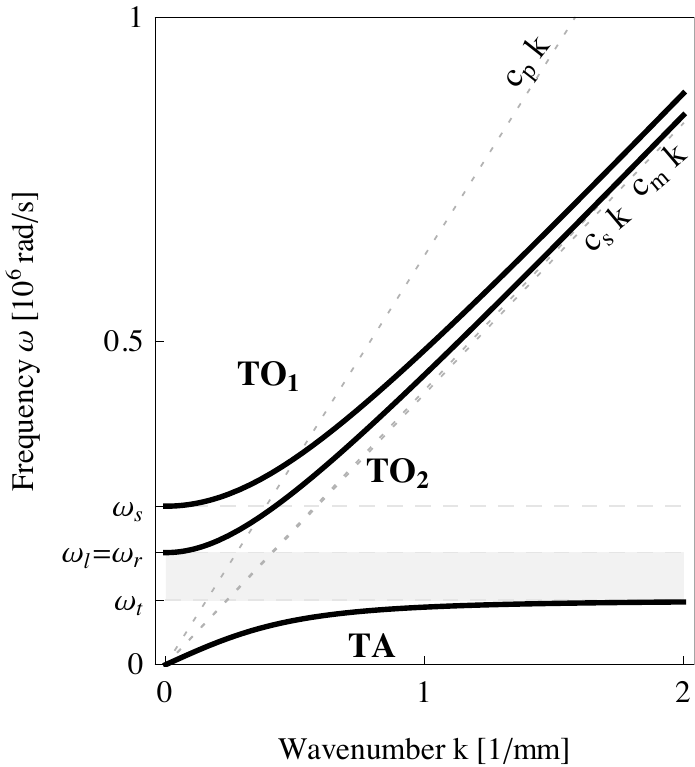} &  & \tabularnewline
(a) & (b) & (c) &  & \tabularnewline
\end{tabular}
\par\end{centering}

\caption{\label{separate0}Dispersion relations $\omega=\omega(k)$ for $\mu_{c}=\mu_{c}^{0}$:
uncoupled waves (a), longitudinal waves (b) and transverse waves (c).
TRO: transverse rotational optic, TSO: transverse shear optic, TCVO:
transverse constant-volume optic, LA: longitudinal acoustic, LO$_{1}$-LO$_{2}$:
first and second longitudinal optic, TA: transverse acoustic, TO$_{1}$-TO$_{2}$:
first and second transverse optic.}
\end{figure}

\begin{figure}[H]
\begin{centering}
\begin{tabular}{ccccc}
\includegraphics[width=0.3\linewidth]{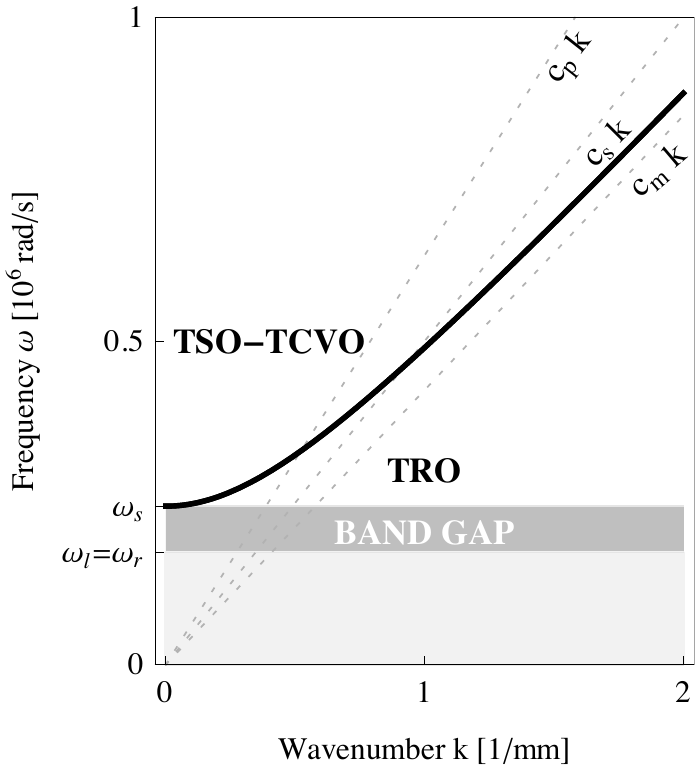}  & \includegraphics[width=0.3\linewidth]{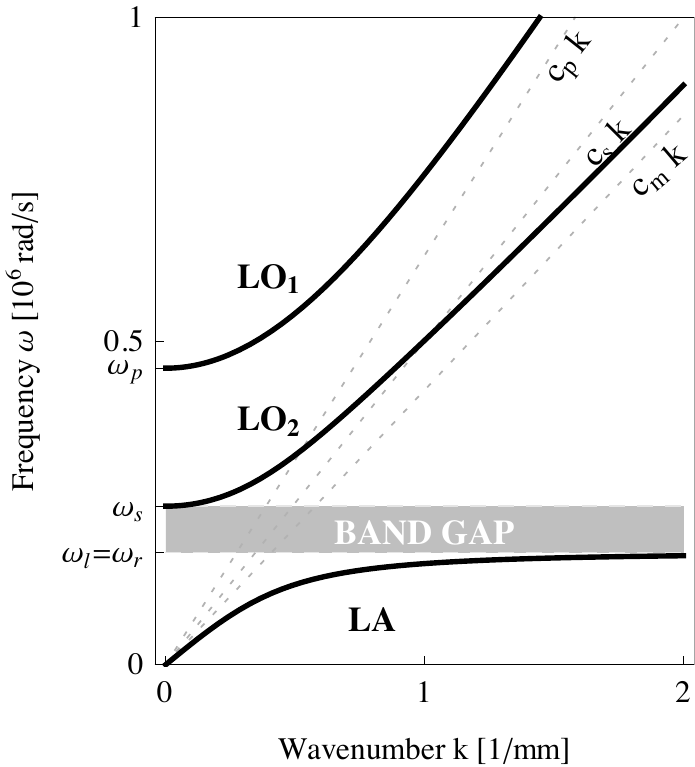} & \includegraphics[width=0.3\linewidth]{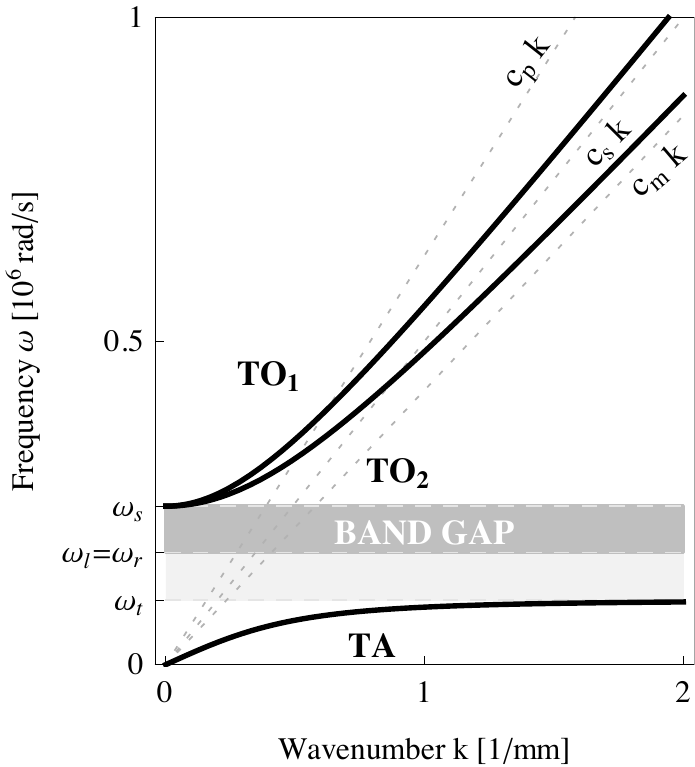} &  & \tabularnewline
(a) & (b) & (c) &  & \tabularnewline
\end{tabular}
\par\end{centering}

\caption{\label{separate}Dispersion relations $\omega=\omega(k)$ for $\mu_{c}=2\:\mu_{c}^{0}$:
uncoupled waves (a), longitudinal waves (b) and transverse waves (c).
TRO: transverse rotational optic, TSO: transverse shear optic, TCVO:
transverse constant-volume optic, LA: longitudinal acoustic, LO$_{1}$-LO$_{2}$:
first and second longitudinal optic, TA: transverse acoustic, TO$_{1}$-TO$_{2}$:
first and second transverse optic.}
\end{figure}

\begin{figure}[H]
\begin{centering}
\begin{tabular}{ccccc}
\includegraphics[width=0.3\linewidth]{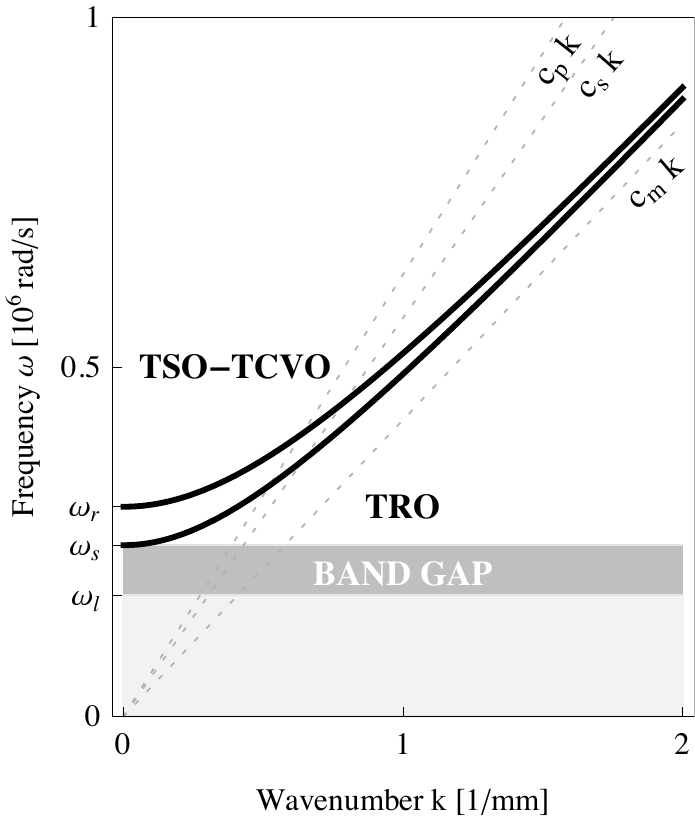}  & \includegraphics[width=0.3\linewidth]{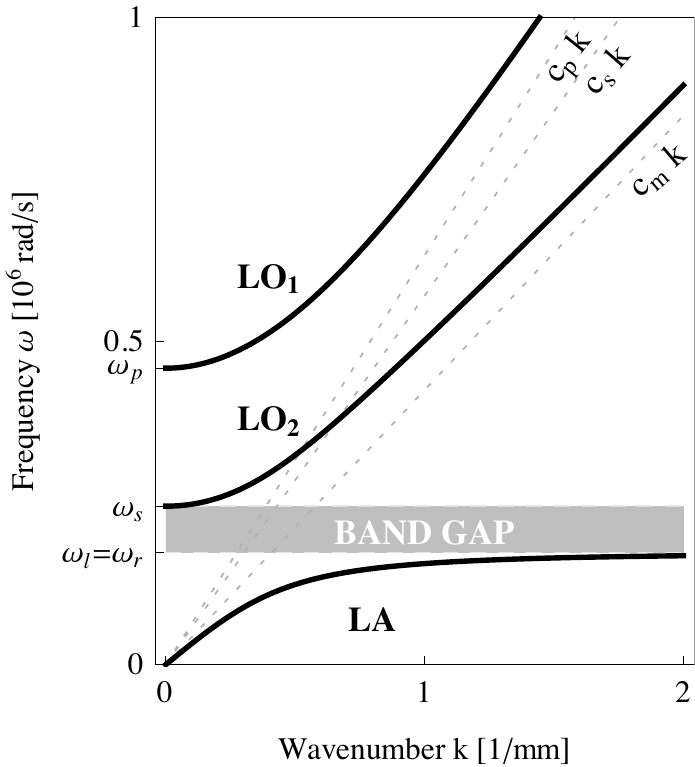} & \includegraphics[width=0.3\linewidth]{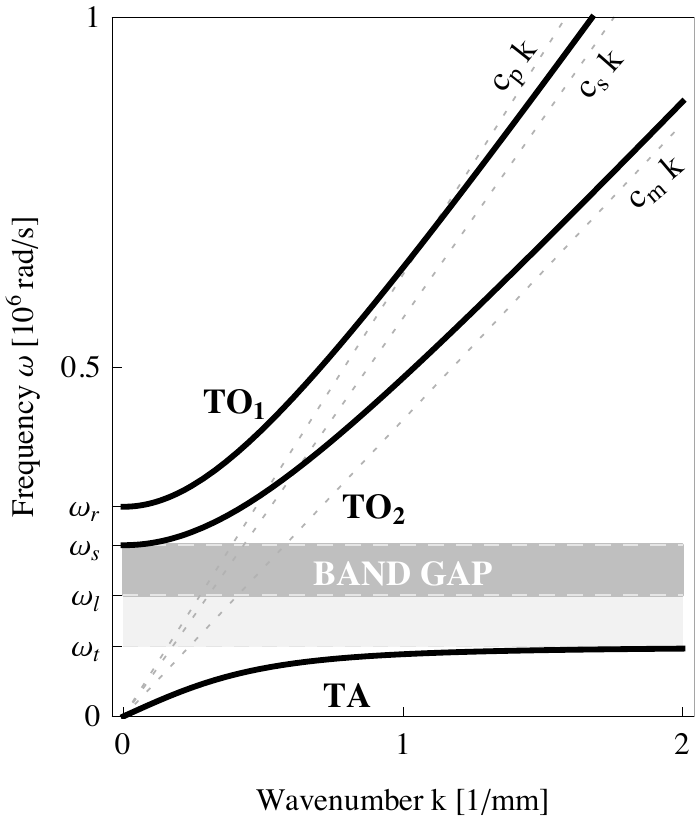} &  & \tabularnewline
(a) & (b) & (c) &  & \tabularnewline
\end{tabular}
\par\end{centering}

\caption{\label{separate1}Dispersion relations $\omega=\omega(k)$ for $\mu_{c}=3\,\mu_{c}^{0}$:
(a), longitudinal waves (b) and transverse waves (c). TRO: transverse
rotational optic, TSO: transverse shear optic, TCVO: transverse constant-volume
optic, LA: longitudinal acoustic, LO$_{1}$-LO$_{2}$: first and second
longitudinal optic, TA: transverse acoustic, TO$_{1}$-TO$_{2}$:
first and second transverse optic.}
\end{figure}

Figure \ref{separate0} shows the dispersion relations for the considered
relaxed micromorphic continuum for the lower bound $\mu_{c}=\mu_{c}^{0}$.
 It can be immediately seen that we cannot identify a frequency band
in which overall wave propagation is forbidden. More precisely, for
any value of frequency, at least on wave exists (longitudinal, transverse
or uncoupled) which propagates inside the considered medium. Things
become different as soon as the value of the Cosserat couple modulus
increases. In Fig. \ref{separate} we show the dispersion relations
for the relaxed micromorphic model corresponding to $\mu_{c}=2\,\mu_{c}^{0}$.
 It can be remarked that in this case any type of wave can propagate
in the frequency range $[\omega_{l},\omega_{s}]$, so that we can
state that a band-gap can be observed which is actually triggered
by the increasing value of the Cosserat couple modulus. In this frequency
range the wavenumber becomes imaginary and only standing waves exist.
For the sake of completeness, we also show in Fig. \ref{separate1}
a third value for the Cosserat couple modulus, namely $\mu_{c}=3\,\mu_{c}^{0}$.
 It can be easily noticed from Fig. \ref{separate1} that the frequency
band gap remains unchanged with respect to the previous case, even
if the global behavior of the system is much more complicated (two
cutoff frequencies for the uncoupled and transverse waves instead
than one). We can hence confirm that the presence of band-gaps is
actually triggered by the Cosserat couple modulus and we can observe
that the wider extension of such band-gaps is reached for $\mu_{c}=2\,\mu_{c}^{0}$.

As shown by Eq. (\ref{eq:bulk-mod-3}), the Cosserat couple modulus
is a homogenized elastic coefficient which allows to account for the
relative rotation of the microscopic inclusions with respect to the
macroscopic matrix. Even if no length scale is associated to this
parameter, the fact that the considered microstructure can rotate
with respect to the matrix in which it is embedded, actually appears
to be a crucial point in order to have the onset of frequency band
gaps at the macroscopic level. In\cite{Neff_ZAMM05} it has been demonstrated
that the Cosserat couple modulus is zero for continuous materials.
Therefore, a positive Cosserat couple modulus accounts for the discreteness
of the considered meta-material.

Wave propagation in complex media exhibiting acoustic and optic branches
generating frequency band-gaps can be compared to wave propagation
in discrete media such as phononic crystals and diatomic chains. The
method proposed in this paper allows to account for a homogenized
behavior of the considered material, so that the forecasted band-gaps
are not directly related to a precise microstructure via a rigorous
homogenization procedure. Indeed, the method proposed in the present
pape, allows to forecast frequency band gaps in precise ranges of
frequency which have been directly associated to particular microstructures
(see e.g. {[}10{]}). For this reason, the macroscopic method presented
here could be of great interest for describing systems with complex
microstructures with a very limited number of constitutive parameters.

It is worth noticing that in our relaxed micromorphic model the observed
forbidden range of frequency is valid for every plane wave solution
(compression and shear). In the papers of Grekova and co-workers \cite{Grekova1,Grekova2}
the reduced Cosserat medium is analyzed. The reduced Cosserat model
can be obtained by our relaxed micromorphic model by setting $\alpha_{c}=0$
and $\mu_{h}\rightarrow\infty$. It is shown that if the reduced Cosserat
model is analyzed, the bulk plane compression wave does not change
compared with the classical case and band-gaps associated to these
waves are not observed, due to the presence of a compressive acoustic
wave. On the other hand, the shear wave is strongly influenced by
the presence of rotational degrees of freedom and a forbidden range
of frequency exists. Seismological observations \cite{Grekova3,Grekova4}
confirm these theoretical investigations and tell us that near earthquake
centers often there are zones absorbing shear waves of a certain band
of frequencies. The micromorphic model proposed in the present paper
can be considered to be more general with respect to other continuum
models which are available in the literature in the sense that global
band gaps (compression and shear) can be forecasted.

For what concerns the complete Cosserat model, it can be obtained
from our relaxed micromorphic model by simply setting $\mu_{h}\rightarrow\infty$.
In this case, no band gaps can be observed at all due to the presence
of an acoustic wave both for the compression and for the shear waves.

We conclude by saying that the relaxed micromorphic model proposed
in \cite{Ghiba,NeffUnif} is able to describe the presence of frequency
band-gaps in which no wave propagation can occur. The presence of
band-gaps is intrinsically related to a critical value of the Cosserat
couple modulus $\mu_{c}$ (see \cite{Neff_JeongMMS08,Jeong_Neff_ZAMM08,Neff_ZAMM05,Neff_Jeong_Conformal_ZAMM08,Neff_Jeong_bounded_stiffness09}
for its interpretation) which must be greater than a threshold value
in order to let band-gaps appear. This parameter can hence be seen
as a discreteness quantifier which starts accounting for lattice discreteness
as soon as it reaches the threshold value specified in Eq.(\ref{Band Gaps}).
This fact is a novel feature of the introduced relaxed model: we claim
that neither the classical micromorphic continuum model nor the Cosserat
and the second gradient ones are able to predict such band-gap phenomena.


\begin{thebibliography}{10}

\bibitem{Piezo1}
U.~Andreaus, F.~dell'Isola, and M.~Porfiri.
\newblock Piezoelectric passive distributed controllers for beam flexural
  vibrations.
\newblock {\em Journal of Vibration and Control}, 10(5):625--659, 2004.

\bibitem{MauginBerezowski}
A.~Berezovski, J.~Engelbrecht, and G.A. Maugin.
\newblock One-dimensional microstructure dynamics.
\newblock {\em Mechanics of Microstructured Solids. Lecture Notes in Applied
  and Computational Mechanics}, 46:21--28, 2009.

\bibitem{isola2}
F.~dell'Isola and S.~Vidoli.
\newblock Continuum modelling of piezoelectromechanical truss beams: An
  application to vibration damping.
\newblock {\em Archive of Applied Mechanics}, 68(1):1--19, 1998.

\bibitem{Economou}
E.N. Economoau and M.~Sigalabs.
\newblock Stop bands for elastic waves in periodic composite materials.
\newblock {\em J. Acoust. Soc. Am.}, 95(4):1734--1740, 1994.

\bibitem{EringenBook}
A.C. Eringen.
\newblock {\em Microcontinuum field theories I. Foundations and Solids}.
\newblock Springer-Verlag, New York, 1999.

\bibitem{Ghiba}
I.D. Ghiba, P.~Neff, A.~Madeo, L.~Placidi, and G.~Rosi.
\newblock The relaxed linear micromorphic continuum: existence, uniqueness and
  continuous dependence in dynamics.
\newblock {\em Mathematics and Mechanics of Solids, arXiv:1308.3762v1
  [math.AP]}, doi: 10.1177/1081286513516972, 2014.

\bibitem{Grekova2}
E.F. Grekova.
\newblock Small perturbations of the spherical prestressed state in a nonlinear
  isotropic elastic reduced {C}osserat medium: waves and instabilities.
\newblock In {\em Proceedings of the International Conference Days on
  Diffraction}, pages 78--82, 2011.

\bibitem{Grekova3}
E.F. Grekova.
\newblock Nonlinear isotropic elastic reduced {C}osserat continuum as a
  possible model for geomedium and geomaterials. spherical prestressed state in
  the semilinear material.
\newblock {\em Journal of Seismology}, 16(4):695--707, 2012.

\bibitem{Neff_JeongMMS08}
J.~Jeong and P.~Neff.
\newblock Existence, uniqueness and stability in linear {C}osserat elasticity
  for weakest curvature conditions.
\newblock {\em Math. Mech. Solids}, 15(1):78--95, 2010.

\bibitem{Jeong_Neff_ZAMM08}
J.~Jeong, H.~Ramezani, I.~M\"unch, and P.~Neff.
\newblock A numerical study for linear isotropic {C}osserat elasticity with
  conformally invariant curvature.
\newblock {\em Z. Angew. Math. Mech.}, 89(7):552--569, 2009.

\bibitem{Kafesaki}
M.~Kafesaki, M.M. Sigalas, and N.~Garc{\'\i}a.
\newblock Frequency modulation in the transmittivity of wave guides in
  elastic-wave band-gap materials.
\newblock {\em Physical Review Letters}, 85(19):4044--4047, 2000.

\bibitem{FrequencyOfBandGaps}
A.~Khelif, B.~Djafari-Rouhani, J.O. Vasseur, and P.A. Deymier.
\newblock Transmission and dispersion relations of perfect and
  defect-containing waveguide structures in phononic band gap materials.
\newblock {\em Physical Review B}, 68:024302, 2003.

\bibitem{Grekova4}
Y.F. Kopnichev and I.N. Sokolova.
\newblock Characteristics of the seismicity and absorption field of s-waves in
  the source region of the {S}umatra earthquake of december 26, 2004.
\newblock {\em Doklady Earth Sciences}, 423(1):1257--1261, 2008.

\bibitem{Grekova1}
M.A. Kulesh, E.F. Grekova, and I.N. Shardakova.
\newblock The problem of surface wave propagation in a reduced {C}osserat
  medium, acoustics of structurally inhomogeneous solid media.
\newblock {\em Geological acoustics, Acoustical Physics}, 55(2):218--226, 2009.

\bibitem{madeo2013wave}
A.~Madeo, P.~Neff, I.D. Ghiba, L.~Placidi, and G.~Rosi.
\newblock Wave propagation in relaxed micromorphic continua: modeling
  metamaterials with frequency band-gaps.
\newblock {\em Continuum Mechanics and Thermodynamics}, doi:
  10.1007/s00161-013-0329-2:1--20, 2014.

\bibitem{MauginBookWaves}
G.~Maugin.
\newblock {\em Nonlinear waves in elastic crystals}.
\newblock Oxford {U}niversity {P}ress, 1999.

\bibitem{isola6}
C.~Maurini, F.~dell'Isola, and J.~Pouget.
\newblock On models of layered piezoelectric beams for passive vibration
  control.
\newblock {\em Journal de Physique}, 115:307--316, 2004.

\bibitem{isola1}
C.~Maurini, J.~Pouget, and F.~dell'Isola.
\newblock Extension of the {E}uler-{B}ernoulli model of piezoelectric laminates
  to include 3d effects via a mixed approach.
\newblock {\em Computers and Structures}, 84(22-23):1438--1458, 2006.

\bibitem{Merkel3}
A.~Merkel and V.~Tournat.
\newblock Dispersion of elastic waves in three-dimensional noncohesive granular
  phononic crystals: Properties of rotational modes.
\newblock {\em Pysical Review E}, 82, 031305, 2010.

\bibitem{Merkel2}
A.~Merkel and V.~Tournat.
\newblock Experimental evidence of rotational elastic waves in granular
  phononic crystals.
\newblock {\em Physical Review Letters}, 107:225502, 2011.

\bibitem{MerkelGranular}
A.~Merkel, V.~Tournat, and V.~Gusev.
\newblock Elastic waves in noncohesive frictionless granular crystals.
\newblock {\em Ultrasonics}, 50:133--138, 2010.

\bibitem{Mindlin63}
R.D. Mindlin.
\newblock Influence of couple-stresses on stress concentrations.
\newblock {\em Experimental Mechanics}, 3(1):1--7, 1963.

\bibitem{Mindlin}
R.D. Mindlin.
\newblock Micro-structure in linear elasticity.
\newblock {\em Arch. Rat. Mech. Analysis}, 16(1):51--78, 1964.

\bibitem{Neff_ZAMM05}
P.~Neff.
\newblock The {C}osserat couple modulus for continuous solids is zero viz the
  linearized {C}auchy-stress tensor is symmetric.
\newblock {\em Z. Angew. Math. Mech.}, 86:892--912, 2006.

\bibitem{Neff_Forest_jel05}
P.~Neff and S.~Forest.
\newblock A geometrically exact micromorphic model for elastic metallic foams
  accounting for affine microstructure. {M}odelling, existence of minimizers,
  identification of moduli and computational results.
\newblock {\em J. Elasticity}, 87:239--276, 2007.

\bibitem{NeffUnif}
P.~Neff, I.D. Ghiba, A.~Madeo, L.~Placidi, and G.~Rosi.
\newblock A unifying perspective: the relaxed linear micromorphic continuum.
\newblock {\em Continuum Mechanics and Thermodynamics}, DOI:
  10.1007/s00161-013-0322-9, 2013.

\bibitem{Neff_Jeong_Conformal_ZAMM08}
P.~Neff and J.~Jeong.
\newblock A new paradigm: the linear isotropic {C}osserat model with
  conformally invariant curvature energy.
\newblock {\em Z. Angew. Math. Mech.}, 89(2):107--122, 2009.

\bibitem{Neff_Jeong_bounded_stiffness09}
P.~Neff, J.~Jeong, and A.~Fischle.
\newblock Stable identification of linear isotropic {C}osserat parameters:
  bounded stiffness in bending and torsion implies conformal invariance of
  curvature.
\newblock {\em Acta Mechanica}, 211(3-4):237--249, 2010.

\bibitem{isola3}
M.~Porfiri, F.~dell'Isola, and E.~Santini.
\newblock Modeling and design of passive electric networks interconnecting
  piezoelectric transducers for distributed vibration control.
\newblock {\em International Journal of Applied Electromagnetics and
  Mechanics}, 21(2):69--87, 2005.

\bibitem{Vasseur1}
J.O. Vasseur, P.A. Deymier, B.~Chenni, B.~Djafari-Rouhani, L.~Dobrzynsk, and
  D.~Prevost.
\newblock Experimental and theoretical evidence for the existence of absolute
  acoustic band gaps in two-dimensional solid phononic crystals.
\newblock {\em Pysical Review Letters,}, 86(14):3012--3015, 2001.

\bibitem{Vasseur}
J.O. Vasseur, P.A Deymier, G.~Frantziskonisx, G.~Hongx, B.~Djafari-Rouhaniy,
  and L.~Dobrzynskiy.
\newblock Experimental evidence for the existence of absolute acoustic band
  gaps in two-dimensional periodic composite media.
\newblock {\em J. Phys. Condens. Matter}, 10:6051, 1998.

\bibitem{Piezo}
S.~Vidoli and F.~dell'Isola.
\newblock Vibration control in plates by uniformly distributed {PZT} actuators
  interconnected via electric networks.
\newblock {\em European Journal of Mechanics, A/Solids}, 20(3):435--456, 2001.

\end{thebibliography}
\end{document}